\documentclass{emulateapj}
\RequirePackage{epstopdf}

\newcommand{\mum}{$\mu$m}
\newcommand{\hoo}{H$_2$O}

\slugcomment{Accepted by ApJL 2012 January 24}

\shorttitle{SOFIA}
\shortauthors{Young et al.}

\begin{document}

\title{Early Science with SOFIA, the Stratospheric Observatory for Infrared Astronomy}

\author{E.T. Young\altaffilmark{1},
E.E. Becklin\altaffilmark{1,2},
P.M. Marcum\altaffilmark{3},
T.L. Roellig\altaffilmark{3},
J.M. De Buizer\altaffilmark{1},
T.L. Herter\altaffilmark{4},
R.G\"{u}sten\altaffilmark{5},
E.W. Dunham\altaffilmark{6},
P. Temi\altaffilmark{3},
B-G Andersson\altaffilmark{1},
D. Backman\altaffilmark{7,8},
M. Burgdorf\altaffilmark{7,9},
L.J. Caroff\altaffilmark{10},
S.C. Casey\altaffilmark{1},
J.A. Davidson\altaffilmark{11},
E.F. Erickson\altaffilmark{10},
R.D. Gehrz\altaffilmark{12},
D.A. Harper\altaffilmark{13},
P.M. Harvey\altaffilmark{14}
L.A. Helton\altaffilmark{1},
S.D. Horner\altaffilmark{3},
C.D. Howard\altaffilmark{1},
R. Klein\altaffilmark{1},
A. Krabbe\altaffilmark{9},
I.S. McLean\altaffilmark{2},
A.W. Meyer\altaffilmark{1},
J.W. Miles\altaffilmark{1},
M.R. Morris\altaffilmark{2},
W.T. Reach\altaffilmark{1},
J. Rho\altaffilmark{7,8},
M.J. Richter\altaffilmark{15},
H-P Roeser\altaffilmark{16},
G. Sandell\altaffilmark{1},
R. Sankrit\altaffilmark{1},
M.L. Savage\altaffilmark{1},
E.C. Smith\altaffilmark{3},
R.Y. Shuping\altaffilmark{1,17},
W.D. Vacca\altaffilmark{1},
J.E. Vaillancourt\altaffilmark{1},
J. Wolf\altaffilmark{7,9},
H. Zinnecker\altaffilmark{7,9}
}

\altaffiltext{1}{SOFIA Science Center, Universities Space Research Association, NASA Ames Research Center, MS 232, Moffett Field, CA 94035, USA}

\altaffiltext{2}{Department of Physics and Astronomy, University of California Los Angeles, 405 Hilgard Avenue, Los Angeles, CA 90095, USA}

\altaffiltext{3}{NASA Ames Research Center, MS 232, Moffett Field, CA 94035, USA}

\altaffiltext{4}{Astronomy Department, 202 Space Sciences Building, Cornell University, Ithaca, NY 14853-6801, USA}

\altaffiltext{5}{Max-Planck Institut f\"{u}r Radioastronomie, Auf dem H\"ugel 69, Bonn, Germany}

\altaffiltext{6}{Lowell Observatory, 1400 W. Mars Hill Rd., Flagstaff AZ 86001, USA}

\altaffiltext{7}{SOFIA Science Center, NASA Ames Research Center, MS 211-1, Moffett Field, CA 94035, USA}

\altaffiltext{8}{SETI Institute, 515 North Whisman Road, Mountain View, CA 94043, USA}

\altaffiltext{9}{Deutsches SOFIA Institut, Universit\"{a}t Stuttgart, Pfaffenwaldring 31, D-70569 Stuttgart, Germany}

\altaffiltext{10}{NASA Ames Research Center, Moffett Field, CA 94035, USA}

\altaffiltext{11}{School of Physics, The University of Western Australia (M013), 35 Stirling Highway, Crawley WA 6009, Australia}

\altaffiltext{12}{Minnesota Institute for Astrophysics, School of Physics and Astronomy, 116 Church Street, S. E., University of Minnesota, Minneapolis, MN 55455, USA}

\altaffiltext{13}{Yerkes Observatory, University of Chicago, 373 W. Geneva St., Williams Bay, WI, USA}

\altaffiltext{14}{Astronomy Department, University of Texas at Austin, 1 University Station C1400, Austin, TX 78712-0259, USA}

\altaffiltext{15}{Department of Physics, University of California at Davis, CA 95616, USA}

\altaffiltext{16}{Institut f\"ur Raumfahrtsysteme, Universit\"{a}t Stuttgart, Pfaffenwaldring 31, D-70569 Stuttgart, Germany}

\altaffiltext{17}{Space Science Institute, 4750 Walnut Street, Boulder, CO 80301, USA}


\begin{abstract}
The Stratospheric Observatory for Infrared Astronomy (SOFIA) is an airborne observatory consisting of a specially modified Boeing 747SP with a 2.7-m telescope, flying at altitudes as high as 13.7\,km (45,000\,ft). Designed to observe at wavelengths from 0.3\,\mum\ to 1.6\,mm, SOFIA operates above 99.8\% of the water vapor that obscures much of the infrared and submillimeter.  SOFIA has seven science instruments under development, including an occultation photometer, near-, mid-, and far-infrared cameras, infrared spectrometers, and heterodyne receivers.  SOFIA, a joint project between NASA and the German Aerospace Center DLR, began initial science flights in 2010 December, and has conducted 30 science flights in the subsequent year.  During this early science period three instruments have flown: the mid-infrared camera FORCAST, the heterodyne spectrometer GREAT, and the occultation photometer HIPO.  This article provides an overview of the observatory and its early performance.

\end{abstract}

\keywords{infrared: general --- instrumentation: miscellaneous --- telescopes}


\section{Introduction}

The Stratospheric Observatory for Infrared Astronomy (SOFIA) is a joint project of the National Aeronautics and Space Administration, USA (NASA) and the German Aerospace Center (DLR).  The NASA and DLR development and operations costs, as well as the science observing time, are divided up in 80:20 proportions, respectively. SOFIA consists of a 2.7-m telescope developed by DLR that resides in a highly modified Boeing 747SP aircraft, enabling observations of a wide variety of astronomical objects at wavelengths from 0.3\,\mum\ to 1.6\,mm \citep{stutzki06, becklin07, gehrz09}.  The SOFIA telescope design and its evolving instrument complement build upon the legacy of NASA's Kuiper Airborne Observatory (KAO), a 0.9\,m infrared telescope that flew from 1974 to 1995 in a Lockheed C141 Starlifter aircraft \citep{gillespie81}.  SOFIA flies at altitudes of up to  13.7\,km (45,000\,ft), which is above 99.8\% of the atmospheric water (\hoo) vapor.  At SOFIA's operational altitudes, the typical precipitable \hoo\ column depth is about 10\,\mum\ (roughly a hundred times lower than at good terrestrial sites).  This enables observations in large parts of the infrared spectrum that are inaccessible from the ground. Figure 1 compares the computed atmospheric transmission from the operating altitude of SOFIA with that from one of the best terrestrial sites, the 5612-m Cerro Chajnantor site in Chile.  The calculations were done using the ATRAN model \citep{lord1992}, and assume a median precipitable water vapor of 0.7 mm for Cerro Chajnantor \citep{radford08}.  In addition to providing access to virtually all of the infrared, SOFIA, with its projected 20-year operational lifetime, serves as a platform for the development of new generations of instruments.  Education and public outreach are important elements of the SOFIA mission.  Consequently, the design of SOFIA included from the beginning provisions for flying educators who would be able to share the scientific experience with students.

A schematic view of the SOFIA facility is shown in Figure 2.  The telescope resides in an open cavity in the aft section of the aircraft and views the sky through a port-side doorway.  The door has a rigid upper segment and a flexible lower segment that can be tracked together to allow the telescope to operate, unvignetted, over an elevation range of 23$\arcdeg$--58$\arcdeg$. The telescope is moved by magnetic torquers around a 1.2-m diameter spherical hydrostatic bearing that floats under an oil pressure of 20 atmospheres within two closely fitting spherical rings.   The rings are mounted in the 6.4-m diameter pressure bulkhead on the axis of the Nasmyth beam. The travel of the bearing for azimuth tracking is only $\pm$3\arcdeg\, so the aircraft heading must be periodically adjusted to keep the source within the telescope field of view. The forward part of the airplane is pressurized, and the working environment for the crew is typical of that in a commercial airliner. This pressurized region provides access to the science instrument during the flight.  The characteristics of the SOFIA observatory are given in Table 1.

The first SOFIA science observations were conducted in 2010 December using the FORCAST (Faint Object InfraRed Camera for the SOFIA Telescope)  mid-infrared camera.  In the eleven months following this first flight, SOFIA conducted 30 science flights using the FORCAST camera, the GREAT (German REceiver for Astronomy at Terahertz frequencies) heterodyne submillimeter spectrometer, and the HIPO (High-speed Imaging Photometer for Occultations) occultation photometer.  This initial phase is known collectively as Early Science and represents a demonstration of initial SOFIA capabilities while the facility and its instrumentation are still under development.

In this paper we provide an overview of the telescope and observatory and describe its performance during the Early Science period.  We set the context for the rest of the articles in this special issue, which present Early Science results based on observations using the FORCAST mid-infrared instrument during the first three flights. An additional 10 flights with FORCAST were conducted for the general astronomical community, a period called Basic Science.  Those results will be described elsewhere.   Observations with HIPO during the Early Science period included an occultation of a star by Pluto on 2011 June 23, during which the observatory was flown through the central chord of the shadow cast by Pluto \citep{person2011}.  The GREAT instrument was flown on 16 scientific flights during Early Science. The first results from those observations will appear in parallel in a special issue of Astronomy and Astrophysics.

\section{SOFIA Telescope and Observatory Overview}

The SOFIA telescope, supplied by DLR as part of the German contribution to SOFIA,  is a bent classical Cassegrain with a 2.690-m diameter parabolic primary mirror, and a 0.352\,m hyperbolic secondary mirror.  The secondary mirror is deliberately undersized and illuminates a 2.5\,m diameter effective aperture to enable chopping without spilling the beam onto the warm region surrounding the primary mirror.  The secondary mirror is attached to a chopping mechanism providing amplitudes of $\pm$4\arcmin\ at chop frequencies up to 20\,Hz. The unvignetted field-of-view (FOV) is 8\arcmin.  The f/19.6 Nasmyth infrared focus is fed by a 45\arcdeg\ gold coated dichroic mirror (Figure 3).  This dichroic mirror separates the light into a reflected infrared beam that is sent to the science instrument and a visible light beam that is sent to a CCD guide camera, the Focal Plane Imager (FPI).    Two other imaging/guiding cameras, the Wide Field Imager (WFI) and the Fine Field Imager (FFI) are used as part of the acquisition system.  They are attached to the front ring of the telescope.  Table 2 gives some of the key optical parameters of the SOFIA telescope and guide cameras.  A more detailed description of the telescope has been given by \citet{krabbe2000}.

The pointing of the telescope is performed more like a space telescope than a ground-based telescope. The primary reference frame for SOFIA is a set of fiber optic gyroscopes that maintain an inertial reference frame.  Once a target is acquired on the sky, its position is stabilized by these  gyroscopes with occasional updates by the guide cameras.

The telescope optics are designed to provide 1.1\arcsec\ FWHM images on-axis at 0.6\,\mum\ ~with diffraction-limited performance at wavelengths longer than 15\,\mum.  However, the telescope is subject to various vibrations as well as variable wind loads in flight, which affect the telescope pointing stability and hence the delivered image quality. SOFIA has active and passive damping systems designed to mitigate some of these effects. In the Early Science flights, after limited tuning of the damping systems, the telescope produced an image quality of 3.2\arcsec\ FWHM at 19.7\,\mum\ with an RMS pointing stability of 1.7$\arcsec$. While these values are large enough to affect image quality at the shorter wavelength end of the SOFIA 0.3--1600~$\mu$m observing range, the observatory was nearly diffraction-limited at $\lambda$$>$38$\mu$m (i.e. over the large majority of the SOFIA operating wavelength range) just after First Light.

One of the larger causes of variable image quality is high velocity turbulent  airflow across the cavity, which drives vibrations and leads to elongation of images in the cross-elevation direction.  Early tests showed that the magnitude of this effect may also depend on the telescope elevation. Therefore, because of variables like the amount of turbulence and cross-elevation vibration, the Point Spread Function (PSF) was not stable in the mid-infrared for Early Science from observation to observation.

Further mitigation of the RMS pointing jitter through tuning of active control systems and the addition of active mass dampers were implemented in late 2011. These dampers are installed at various locations on the primary and secondary mirror support structures. In the future, after optimization of the damping systems, the RMS pointing stability is expected to improve to 0.5\arcsec, which will deliver an image quality of 2.1\arcsec\ FWHM at 19.7\,\mum.

At wavelengths shorter than 10\,\mum, another significant contributor to the point spread function width is the seeing due to turbulence in the shear layer in the vicinity of the airplane.  The shear layer seeing is wavelength dependent and increases at shorter wavelengths.  Below roughly 5\,\mum, the shear layer seeing becomes the dominant image size contributor.

The secondary mirror is capable of chopping in any direction from center with a maximum amplitude of 292\arcsec, with the additional constraint of a maximum total chop amplitude of 480\arcsec.  As is the case for all Cassegrain systems, tilting the secondary mirror produces coma.  For the SOFIA telescope, the coma causes point spread function smearing in the direction of the chop by 2\arcsec   ~per 1\arcmin  ~ of chop amplitude.

\section{SOFIA Instrumentation}

One of the strengths of the SOFIA observatory is the ability to change instruments.  Seven first-generation instruments have been developed or are being developed for SOFIA. Collectively they span a large wavelength range (0.3--250\,\mum ) and include imagers and spectrographs.  SOFIA provides several classes of instruments.  Facility Class instruments are general purpose instruments that are maintained and operated by the observatory.  Principal Investigator Class instruments are maintained and supported by the instrument teams.  Special Purpose instruments are specialized capabilities also supported by the instrument teams.

FORCAST (Principal Investigator, PI, Terry Herter, Cornell University) is a mid-infrared camera operating between 5 and 40~\mum. The instrument is described in the following paper by \citet{herter2012}.
GREAT  (PI Rolf G\"{u}sten, MPIfR)\citep{heyminck2012} is a heterodyne spectrometer currently operating in the 1.2 - 2.5 THz frequency range.  These instruments were used during Early Science for both guaranteed time observations planned by the instrument teams and for programs from the astronomical community competed via open proposal calls.

Two other instruments, HIPO (High-speed Imaging Photometer for Occultations) (PI Edward Dunham, Lowell Observatory)\citep{dunham2004} and FLITECAM (First Light Infrared Test Experiment CAMera, PI Ian McLean, UCLA) \citep{mclean2006} were used on development flights on SOFIA.  HIPO consists of a pair of high-speed CCD detectors specifically configured to be used for occultation and other high speed imaging applications.  FLITECAM is a near-infrared camera operating between 1 and 5~\mum.  The instrument also has grisms installed that provide moderate resolution spectroscopy at these wavelengths.  All four instruments will be offered to the astronomy community for the first annual observing cycle (Cycle 1).

The remaining three instruments, EXES (Echelon-Cross-Echelle Spectrograph, PI Matt Richter, UC Davis)~\citep{richter2003}, FIFI-LS (Field Imaging Far-Infrared Line Spectrometer, PI Alfred Krabbe, University of Stuttgart) \citep{klein2010} and HAWC (High-resolution Airborne Wideband Camera, PI Doyal Harper, U. Chicago) \citep{harper2004} are expected to be offered in a second annual Call for Proposals (Cycle 2).  Table 4 summarizes the characteristics of the First Generation SOFIA Instruments.  The SOFIA program also plans to introduce new instruments in future years.

\section{SOFIA Operations}

SOFIA is based at the NASA Dryden Aircraft Operations Facility (DAOF) in Palmdale, California.  The DAOF is a multi-user NASA facility that supports a number of scientific aircraft.  The SOFIA Operations Center (SOC) is located at the DAOF and performs the mission operations and laboratory support for the observatory.  The SOFIA Science Center (SSC) is located at the NASA Ames Research Center in Moffett Field, California.  Collectively, the SSC and the SOC combine to make up SOFIA Science Mission Operations (SMO), which is responsible for the conduct of the science on SOFIA. The overall SOFIA mission development and operations are managed by NASA's SOFIA Program Office, currently located at the DAOF.  The SOFIA aircraft operations are managed by NASA's Dryden Flight Research Center (DFRC), which provides the aircraft maintenance staff, aircraft safety personnel, the pilots, navigators, and other aircraft flight crew members.
The SMO is jointly managed by the Universities Space Research Association (USRA) for NASA and by the Deutsches SOFIA Institut (DSI), in Stuttgart, Germany for DLR\@.   Science support for the user community will be provided by the SMO and the DSI\@.

Typical crew members during the Early Science period consisted of the flight deck (pilot, co-pilot, navigator, and flight engineer), Mission Operations (mission director, science flight planner, telescope operators), instrument team, engineering support (safety technicians, telescope engineers, and water vapor monitor engineer), and General Investigators.

SOFIA will mainly operate from its home base at DAOF but will also be deployed to operate from other bases in various parts of the world.  In particular, regular deployments to the Southern Hemisphere are planned to provide access to targets that cannot be observed from the Palmdale base.

To maximize the operational efficiency of SOFIA, observations are all queue scheduled.  Starting in Cycle 1, the bulk of SOFIA observing time will be awarded on a yearly basis via open, peer-reviewed proposal calls, and observations will be awarded in units of time.  Thus there will be a process of flight-planning for every observing cycle which will be very much like planning year-long observations for a space mission. For SOFIA there will be the added constraints imposed by the limited sky-visibility determined by aircraft heading.  Also, sensitive observations requiring the driest conditions need to be scheduled for the highest possible aircraft altitudes. After take-off, initial aircraft cruising altitude is 11.6-11.9 km (38000-39000~ft). As fuel is burned off, the aircraft can fly higher. The performance of the aircraft allows for 6 hours of observing time above 12.5 km (41000~ft) and of which 4 hours can be above 13.1 km (43000~ft).  The maximum flight time under routine operations is 10 hours.

The flight planning process involves selecting suitable targets from a larger pool that satisfy requirements such as correct azimuth, necessary altitude, and instrument configurations.  To ensure efficient use of observing time, the SOFIA Flight Planning team needs to have available a pool of targets well distributed on the sky.  Figure 4 illustrates a typical flight plan from the FORCAST Basic Science period.  The flight begins in Palmdale and the plan includes a mixture of galactic and extragalactic targets as well as northern and southern targets.  The requirement to return to Palmdale combined with a nominal 10-hour flight length implies a maximum leg length of approximately 4 hours on a given source.

During the 11 months of Early Science, SOFIA conducted 30 observational flights with more than 200 hours of research time.  SOFIA is planned to ultimately fly approximately 120 flights per year with more than 960 research hours.

\section{Accessing Early Science data from the SOFIA Archive}

Data accumulated during all flights are archived in the SOFIA Data Archive. The raw data (Level 1) obtained during science flights are ingested into the archive and made available for download to the program General Investigator (GI) within 24 hours. These data are available in standard FITS format, with header information and appropriate Flexible Image Transport System (FITS) keywords conforming to the SOFIA FITS keyword dictionary.  Raw FORCAST data must be corrected for several instrumental effects, including response non-linearity, field distortion, and image artifacts. SOFIA Mission Operations is responsible for this processing of the FORCAST data via a software pipeline. These Level 2 data (corrected for instrumental and atmospheric effects) are scientifically valid and available for download from the SOFIA Archive within two weeks after the completion of the flight series in which they are taken.

As a Principal Investigator class instrument, the GREAT data follow a somewhat different path.  The raw data are archived in the SOFIA Data Archive, but the instrument team provides support for the subsequent processing of the data into scientifically useful products.

Nominally, data is accessible to the general community after a proprietary period of 12 months, starting from the time of ingestion into the archive. However, FORCAST Early Science data, including those presented in this volume, have a limited proprietary period and are now available to the community. Readers wishing to download FORCAST Early Science data must register an account with the SOFIA Data Cycle System (DCS), linked to and accessible from the SOFIA World Wide Web home page (https://dcs.sofia.usra.edu).

\section{Concluding Remarks}

SOFIA has completed over 30 science flights in the year since the initial Early Science Flight in 2010 December.  Some of the results of the first four flights with FORCAST are presented here in this issue, with many more results from Early Science in preparation.  Another set of papers based on GREAT observations is being published in parallel in Astronomy \& Astrophysics.  With the range of current and future instrumentation, and with the projected 20 year operational lifetime, it is expected that SOFIA observations will yield a rich collection of scientific results of which these are the preview.

\acknowledgements

SOFIA science mission operations are conducted jointly by the Universities Space Research Association, Inc. (USRA), under NASA contract NAS2-97001, and the Deutsches SOFIA Institut (DSI) under DLR contract 50 OK 0901 to the University of Stuttgart.  RDG was supported by NASA and the United States Air Force.


\clearpage
\begin{deluxetable}{ll}
\tablecaption{SOFIA Observatory Characteristics}
\tablewidth{0pt}
\tablehead{
\colhead{Characteristic} & \colhead{Value}
}

\startdata

Operational Wavelength & 0.3 -- 1600\,\mum \\
Clear aperture diameter & 2.5\,m \\
Telescope elevation range (unvignetted) & 23\arcdeg\ -- 58\arcdeg \\
Unvignetted field-of-view & 8\arcmin \\
Telescope optics image quality at 0.6\,\mum \tablenotemark{a}
          & 1.6\arcsec\ (80\% encircled energy) \\
Diffraction limited image size (FWHM)\tablenotemark{b}
          & 0.0825 $\times\ \lambda(\mu$m) (\arcsec)\\
Chopper frequencies (2-pt square wave) & 1--20\ Hz  \\
Maximum chop throw (unvignetted) & $\pm 5$\arcmin \\
Maximum flight altitude & 13.7~km (45000~ft) \\
Typical flight duration & 10~h \\
Flight time above 12.5 km (41000~ft) & 6~h \\
Operational Lifetime & 20 years \\
\enddata

\tablenotetext{a}{Does not include the effects of seeing or pointing stability, which dominate at the shorter wavelengths.}
\tablenotetext{b}{The goal at full operational capability is diffraction limited imaging for $\lambda > 15 \mu$m.}

\end{deluxetable}


\begin{deluxetable}{ll}
\tablecaption{SOFIA Telescope Parameters}
\tablewidth{0pt}
\tablehead{
\colhead{Parameter} & \colhead{Value}
}

\startdata

Optical Configuration & Classical Bent Cassegrain \\
Primary mirror diameter & 2.69\,m \\
Primary mirror material & Zerodur \\
Primary conic constant & -1 \\
Primary focal length & 3200 mm \\
Secondary mirror diameter & 352 mm \\
Secondary mirror material & Silicon carbide \\
Secondary radius of curvature & 954.13 mm \\
Secondary conic constant & -1.2980 \\ [10pt]

Effective entrance pupil diameter & 2.5\,m \\
Nominal focal length & 49141 mm \\
Nominal system f-ratio & f/19.6 \\ [10 pt]

Wide Field Imager Field of View &  6\arcdeg x 6\arcdeg  \\
Fine Field Imager Field of View &  70\arcmin x 70\arcmin   \\
Focal Plane Imager Field of View &   8\arcmin x 8\arcmin  \\

\enddata

\end{deluxetable}


\begin{deluxetable}{llllcc}
\tablecaption{SOFIA First Generation Instruments}
\tablewidth{0pt}
\tabletypesize{\scriptsize}
\tablehead{
\colhead{Name} & \colhead{Description} & \colhead{PI} & \colhead{Institution} &\colhead{Wavelengths} & \colhead{Spectral}\\
\colhead{} & \colhead{} & \colhead{} & \colhead{} & \colhead{(\mum)} & \colhead{Resolution}
}

\startdata

FORCAST & Mid Infrared Camera and Grism Spectrometer & T. Herter & Cornell & 5 - 40 & 200 \\
GREAT & Heterodyne Spectrometer & R. G\"{u}sten & MPIfR & 60 - 240 & $10^6 - 10^8$\\
FLITECAM & Near Infrared Camera and Grism Spectrometer & I. McLean & UCLA & 1 - 5 & 2000 \\
HIPO & CCD Occultation Photometer & T. Dunham  & Lowell Obs & 0.3 - 1.1 \\
EXES & Mid-Infrared Spectrometer & M. Richter  & UC Davis & 5 - 28 & $3000, 10^4 , 10^5 $ \\
HAWC & Far-Infrared Camera & D.A. Harper & U Chicago & 50 - 240\\
FIFI-LS & Integral Field Far Infrared Spectrometer & A. Krabbe & U Stuttgart & 42 - 210 & 1000 - 3750 \\

\tablenotetext{a}{Details available at - http://www.sofia.usra.edu/Science/instruments}

\enddata

\end{deluxetable}

\clearpage
\begin{center}

\includegraphics[width=7.0in]{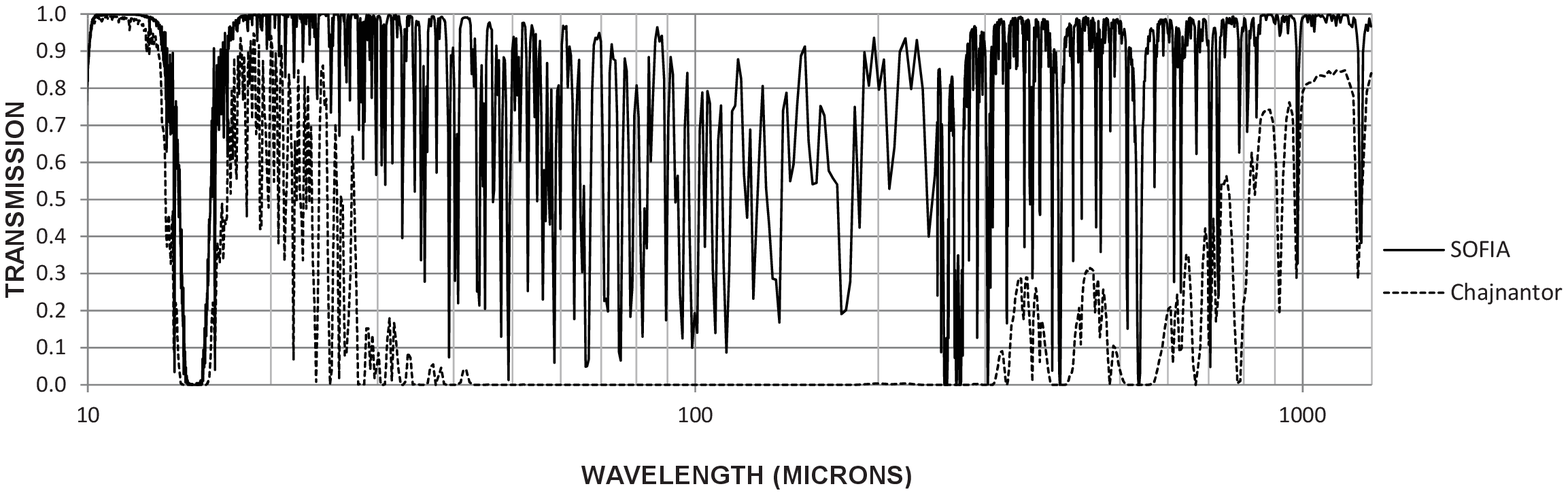}
\figcaption{Calculated atmospheric transmission for SOFIA and Cerro Chajnantor.  Computed using the ATRAN program \citep{lord1992} assuming 10~\mum~and 700~\mum~of precipitable water vapor, respectively.  The average SOFIA transmission between 20~\mum~and 1.3 mm is 80\%}

\end{center}

\begin{center}

\includegraphics[width=7.0in]{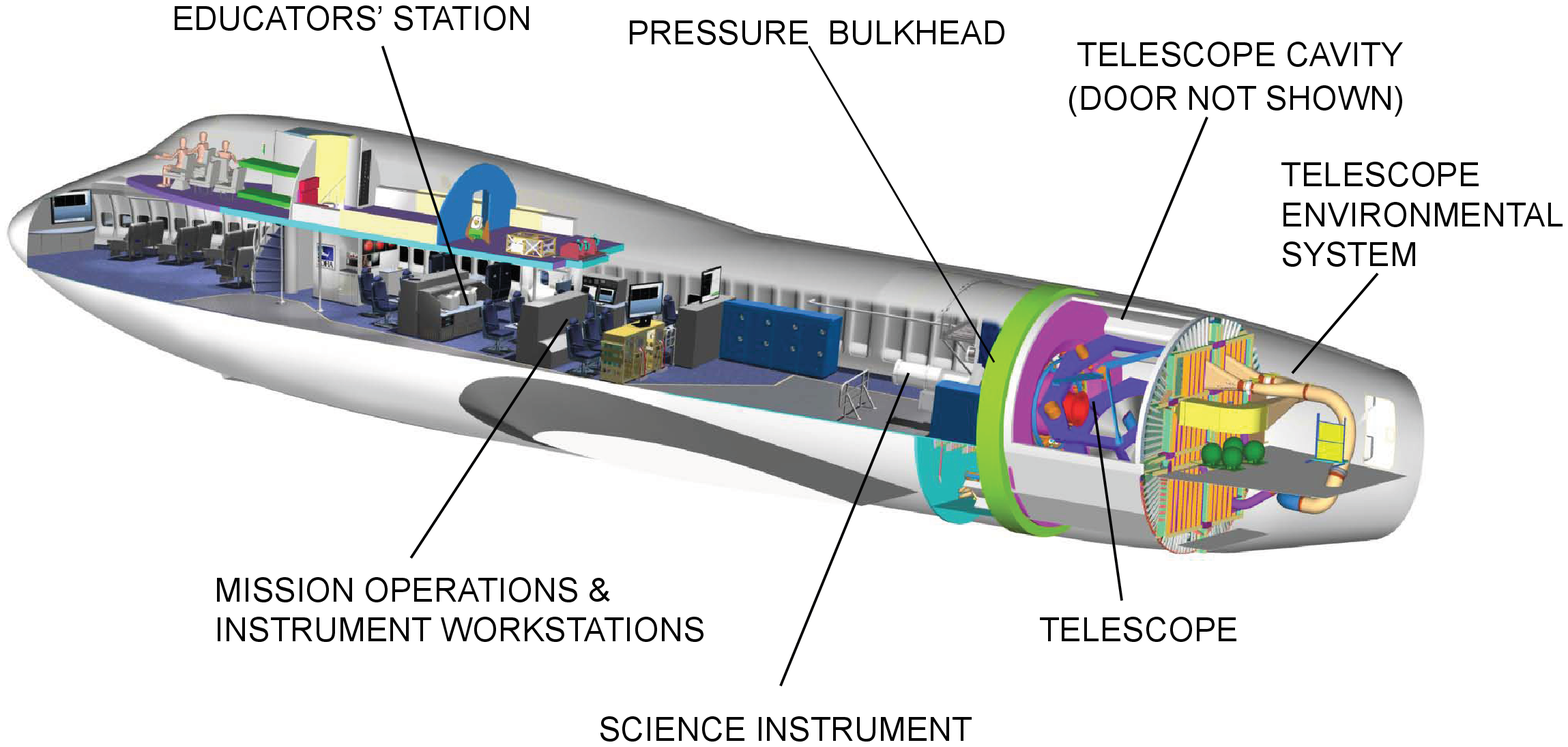}
\figcaption{A cutaway schematic of the SOFIA observatory.  The pressurized cabin is to the left of the pressure bulkhead.  The telescope looks out the port side of the airplane.  The 10000 kg mass is supported by a hydrostatic spherical bearing that allows an elevation range of $23\arcdeg$ to $58\arcdeg$.  The light for the science instrument is fed through the Nasmyth tube which goes through the bearing.  An environmental control system prevents condensation during descent when the door is closed after a night of observations.}

\end{center}

\begin{center}

\includegraphics[width=7.0in]{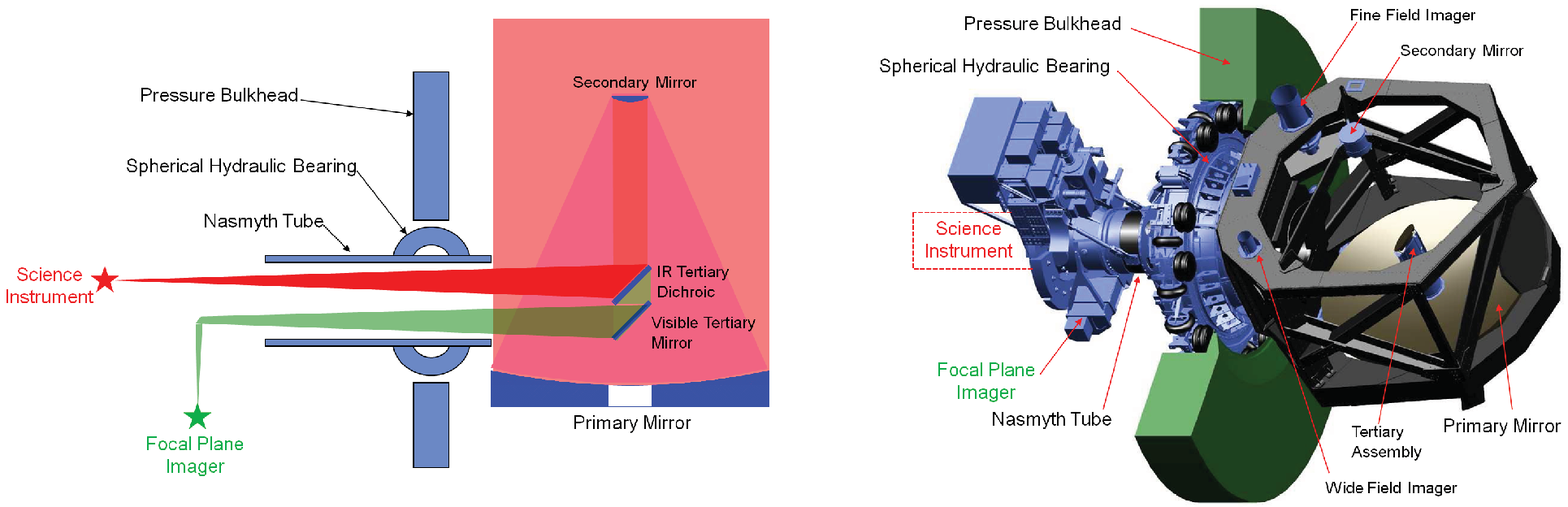}
\figcaption{A schematic of the optical system of the SOFIA telescope (left) and a model of the SOFIA telescope assembly (right). Everything left of the bulkhead is contained in the forward pressurized crew cabin, while everything to the right is contained in the open telescope cavity.}

\end{center}
\begin{center}

\includegraphics[width=3.5in]{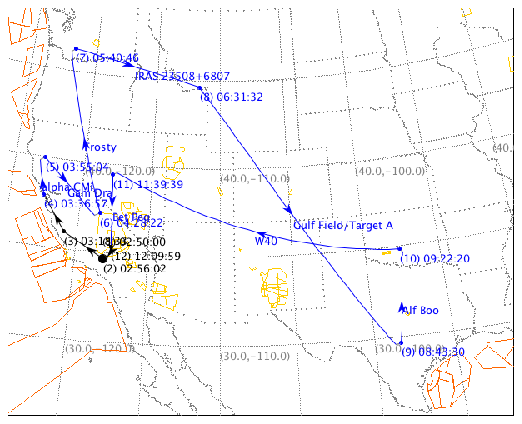}
\figcaption{A representative SOFIA flight plan for the FORCAST Basic Science period. Targets came from the pool of selected General Investigator proposals. Total flight duration for this plan is 9.5 hours. }

\end{center}

\end{document}